\documentclass[11pt, onecolumn]{article}

\usepackage{graphicx}
\usepackage{amsmath}
\usepackage{amssymb}
\usepackage{geometry}
\usepackage{authblk}
\usepackage{hyperref}
\usepackage{caption}
\usepackage{subcaption}
\usepackage{booktabs}
\usepackage[superscript]{cite}
\usepackage{tikz}
\usetikzlibrary{calc, positioning, arrows.meta, decorations.pathreplacing}

\newcommand\blfootnote[1]{%
  \begingroup
  \renewcommand\thefootnote{}\footnote{#1}%
  \addtocounter{footnote}{-1}%
  \endgroup
}

\title{\textbf{\Large Geometry Challenges Entropy: Regime-Dependent Rectification in Nanofluidic Cascades}\\[0.4em]\large Particle size tunes the transition between boundary-dominated accumulation and entropic funnel rectification}
\author{Ting PENG$^{1\ast}$}
\date{}

\begin{document}

\maketitle

\noindent\textit{This manuscript is currently under peer review at Nature Physics (submitted 4 February 2026).}

\vspace{0.4cm}

\noindent\textbf{Can geometry alone reshape equilibrium?} Cascaded nanofluidic chambers show complex accumulation patterns, traditionally attributed to geometric diode effects. We use 3D molecular dynamics to decouple funnel rectification from boundary reflection. Simulations with argon parameters ($r = 0.19$\,nm) reveal a striking ``reverse'' rectification in a 2-chamber setup: the narrow side accumulates over 5$\times$ more particles ($N_1/N_0 = 5.37 \pm 0.01$, $p < 0.0001$). In a 10-chamber argon cascade, this effect drives massive downstream accumulation. A symmetric control ($w_{\mathrm{L}}=w_{\mathrm{R}}$) eliminates the gradient, confirming that funnel asymmetry---not boundary/edge effects---is the primary driver in the ballistic regime. By contrast, the super-atom regime is dominated by boundary reflection. \textbf{Our results challenge standard entropic transport theory and provide design rules for passive, geometry-driven density gradients---no pump, no drive.}

\blfootnote{$^1$Key Laboratory for Special Area Highway Engineering of Ministry of Education, Chang'an University, Xi'an 710064, China. $^\ast$Corresponding author: Ting Peng (e-mail: t.peng@ieee.org). ORCID: \url{https://orcid.org/0009-0001-9059-2278}}


\noindent\textbf{Subject areas:} Statistical physics, Nanofluidics, Molecular dynamics.

\vspace{0.5cm}

The Second Law dictates that isolated systems tend toward maximum entropy---uniform distribution. Yet geometry can reshape equilibrium. Transport in confined nanofluidic structures \cite{schoch2008,holt2006,majumder2005,bocquet2021} exhibits surface-dominated phenomena; cascaded chambers show end accumulation and middle depletion. The prevailing view attributes such effects to funnel asymmetry (geometric diodes) \cite{reguera2006,hanggi2009}. We ask: is it the funnel or the cascade? Using symmetric controls and a dedicated 2-chamber funnel-isolation experiment, we provide the first quantitative demonstration that end accumulation in cascaded nanochannels arises from \emph{reflection at system boundaries}---not funnel asymmetry. Asymmetric ($w_{\mathrm{L}} > w_{\mathrm{R}}$) and symmetric ($w_{\mathrm{L}}=w_{\mathrm{R}}$) 3D hard-sphere MD, plus 2-chamber runs with argon parameters, separate the two effects: boundary reflection is universal; funnel rectification is regime-dependent. The result matters for passive separations, osmotic energy harvesting, and nanoscale Brownian devices---all without applied fields.

\section*{Geometry and Mechanism}

\subsection*{Spatial Configuration}
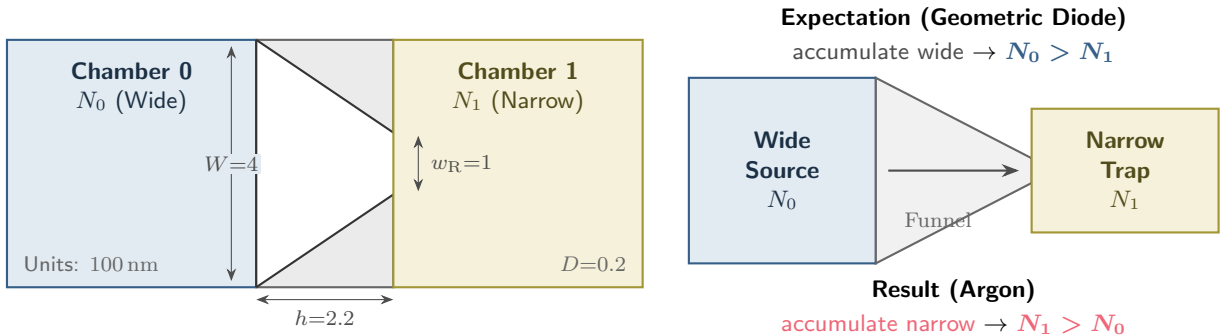
\begin{figure}[t]
    \centering
    \definecolor{nblue}{RGB}{68,119,170}
    \definecolor{ncyan}{RGB}{102,204,238}
    \definecolor{ngreen}{RGB}{34,136,51}
    \definecolor{nyellow}{RGB}{204,187,68}
    \definecolor{nred}{RGB}{238,102,119}
    \definecolor{ngray}{RGB}{187,187,187}

    \begin{subfigure}[c]{0.56\textwidth}
        \centering
        \begin{tikzpicture}[x=0.82cm, y=0.82cm, font=\sffamily\footnotesize, >=Stealth]
            \draw[fill=nblue!15, thick, draw=nblue!80!black] (0,0) rectangle (4,4);
            \node[nblue!40!black] at (2, 3.5) {\bfseries Chamber 0};
            \node[nblue!40!black] at (2, 3.0) {$N_0$ (Wide)};

            \draw[fill=ngray!30, thick, draw=black!60] (4,0) -- (6.2,0) -- (6.2,4) -- (4,4) -- cycle;
            \fill[white] (4,0) -- (4,4) -- (6.2,2.5) -- (6.2,1.5) -- cycle;
            \draw[thick, draw=black!80] (4,0) -- (4,4);
            \draw[thick, draw=black!80] (4,4) -- (6.2,2.5);
            \draw[thick, draw=black!80] (6.2,1.5) -- (4,0);

            \draw[fill=nyellow!20, thick, draw=nyellow!80!black] (6.2,0) rectangle (10.2,4);
            \node[nyellow!40!black] at (8.2, 3.5) {\bfseries Chamber 1};
            \node[nyellow!40!black] at (8.2, 3.0) {$N_1$ (Narrow)};

            \draw[<->, thin, black!70] (3.6, 0.1) -- (3.6, 3.9) node[midway, fill=nblue!15, inner sep=1pt, text=black!70] {\scriptsize $W{=}4$};
            \draw[<->, thin, black!70] (6.6, 1.55) -- (6.6, 2.45) node[midway, right=1pt, text=black!70] {\scriptsize $w_{\mathrm{R}}{=}1$};
            \draw[<->, thin, black!70] (4.0, -0.2) -- (6.2, -0.2) node[midway, below, text=black!70] {\scriptsize $h{=}2.2$};

            \node[text=black!60, anchor=south west] at (0.1, 0.1) {\scriptsize Units: $100\,\mathrm{nm}$};
            \node[text=black!60, anchor=south east] at (10.1, 0.1) {\scriptsize $D{=}0.2$};
        \end{tikzpicture}
        \caption{\textbf{Simulation Unit Cell (3D).} Geometry parameters.}
        \label{fig:geometry_sim}
    \end{subfigure}
    \hfill
    \begin{subfigure}[c]{0.42\textwidth}
        \centering
        \begin{tikzpicture}[x=0.82cm, y=0.82cm, font=\sffamily\footnotesize, >=Stealth]
            \draw[fill=nblue!15, thick, draw=nblue!80!black] (0,0.5) rectangle (3,3.5);
            \node[align=center, nblue!40!black] at (1.5, 2.0) {\bfseries Wide\\\bfseries Source\\$N_0$};

            \draw[thick, fill=ngray!20, draw=black!60] (3,0.5) -- (3,3.5) -- (5.5,2.2) -- (5.5,1.8) -- cycle;
            \draw[->, thick, black!70] (3.2, 2.0) -- (5.3, 2.0);
            \node[text=black!60, font=\scriptsize] at (4.0, 1.2) {Funnel};

            \draw[fill=nyellow!20, thick, draw=nyellow!80!black] (5.5,1.0) rectangle (8.5,3.0);
            \node[align=center, nyellow!40!black] at (7.0, 2.0) {\bfseries Narrow\\\bfseries Trap\\$N_1$};

            \node[align=center] at (4.25, 4.2) {%
                \textbf{\footnotesize Expectation (Geometric Diode)}\\[0.2em]
                \textcolor{black!70}{accumulate wide} $\to$ \textcolor{nblue!80!black}{\boldmath$N_0 > N_1$}
            };

            \node[align=center] at (4.25, -0.2) {%
                \textbf{\footnotesize Result (Argon)}\\[0.2em]
                \textcolor{nred}{accumulate narrow} $\to$ \textcolor{nred}{\boldmath$N_1 > N_0$}
            };
        \end{tikzpicture}
        \caption{\textbf{Mechanism Schematic.} Contrast between heuristic and result.}
        \label{fig:schematic_sim}
    \end{subfigure}
    \caption{\textbf{System Geometry and Mechanism.} (a) Top-down view of the 3D simulation unit cell ($W{\times}H{=}4{\times}4$, $h{=}2.2$). Particles flow between wide ($w_{\mathrm{L}}{=}4$) and narrow ($w_{\mathrm{R}}{=}1$) openings. (b) Schematic comparison: Classical geometric diode theory predicts accumulation in the wide chamber ($N_0 > N_1$) due to funnel reflection. Our argon simulations reveal the opposite: massive accumulation in the narrow chamber ($N_1 > N_0$), driven by regime-dependent funnel rectification.}
    \label{fig:geometry_combined}
\end{figure}

Consider a 3D nanofluidic channel $W \times H \times D$, divided by partitions of thickness $h$. Each partition contains a square-base pyramid funnel opening: wide $w_{\mathrm{L}}$ (source) and narrow $w_{\mathrm{R}}$ (trap). Figure~\ref{fig:geometry_combined} shows the two-chamber unit cell: $W = H = 4$, $h = 2.2$, $w_{\mathrm{L}} = 4$, $w_{\mathrm{R}} = 1$, slab depth $D = 0.2$ (units: 1 = 100\,nm).

\subsection*{Flux Balance}
At equilibrium, $J_{L\to R} = J_{R\to L}$. The entropic barrier makes transmission from wide to narrow harder than narrow to wide---particles from the wide side collide with converging funnel walls and are reflected. Thus $n_L > n_R$: the wide (source) side accumulates particles to balance the reduced flux. 2-chamber runs with argon parameters (below) show the opposite ($N_1 > N_0$), indicating regime dependence \cite{reguera2006,hanggi2009}.

\section*{Results}

\subsection*{Cascaded Amplification and Scaling}
\begin{figure}[t]
    \centering
    \begin{subfigure}[b]{0.48\textwidth}
        \centering
        \includegraphics[width=\textwidth]{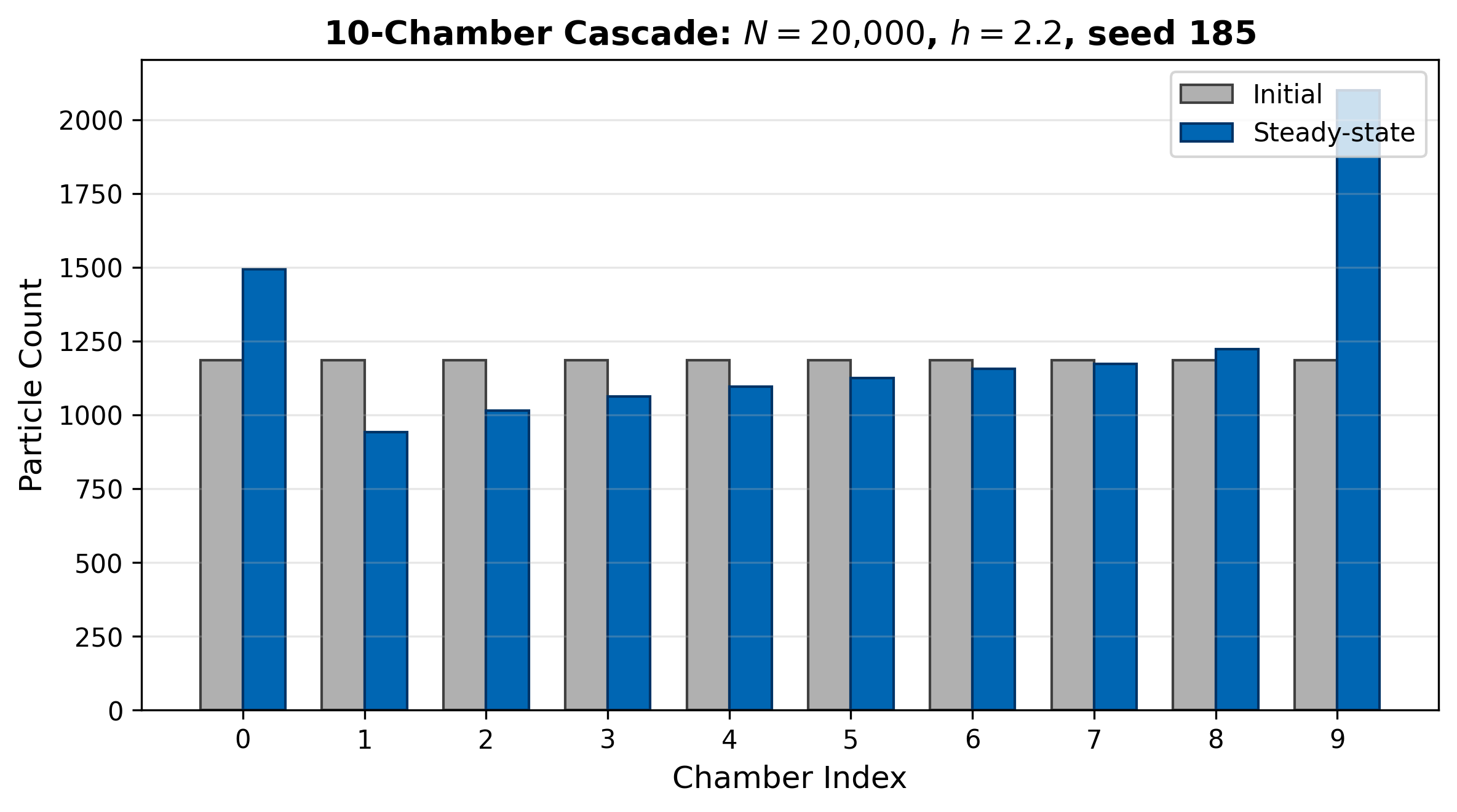}
        \caption{10 chambers: $N = 20{,}000$.}
        \label{fig:cascade_10chambers}
    \end{subfigure}
    \hfill
    \begin{subfigure}[b]{0.48\textwidth}
        \centering
        \includegraphics[width=\textwidth]{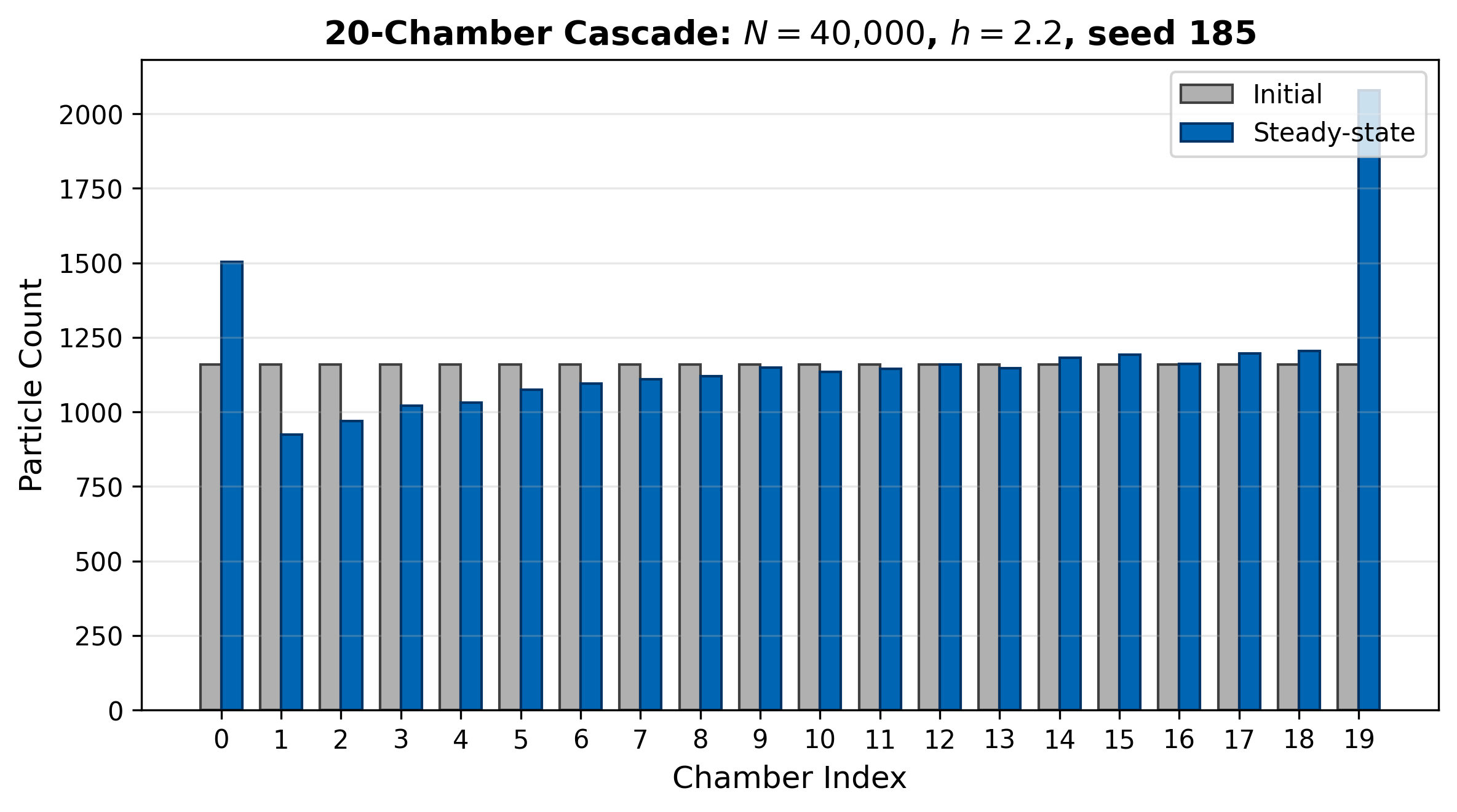}
        \caption{20 chambers: $N = 40{,}000$.}
        \label{fig:cascade_20chambers}
    \end{subfigure}
    \caption{\textbf{Cascaded amplification (seed 185).} Initial uniform vs steady-state. Boundary reflection drives end accumulation; gradient sharpens with chain length. Funnel asymmetry adds rectification on top.}
    \label{fig:cascade}
\end{figure}

We linked 10 and 20 chambers in series, initializing with uniform density (volume-proportional). Figure~\ref{fig:cascade} shows particles accumulating at the ends (with significantly higher accumulation in the final chamber than in Chamber~0) and depleting in the middle. The accumulation pattern persists across different chain lengths. End chambers exchange particles with only one neighbor, whereas middle chambers exchange with two. This connectivity difference drives accumulation at the boundaries. As we show below, a symmetric control demonstrates that this pattern arises from boundary reflection---not from funnel asymmetry.

\subsection*{Validation with Argon Parameters}
To confirm the effect with physically accurate parameters, we ran the 10-chamber cascade using the van der Waals radius of argon ($r = 0.0019$, 0.19\,nm) in simulation: $N = 100{,}000$, $h = 2.2$, 16 seeds in parallel. Figure~\ref{fig:argon_cascade} and Table~\ref{tab:asym_vs_sym} show a striking downstream accumulation: particles pump preferentially from wide to narrow, leading to a massive buildup at the end of the chain ($N_9 \approx 32{,}900$ vs $N_0 \approx 8{,}100$). This ``reverse diode'' effect ($N_{\mathrm{narrow}} > N_{\mathrm{wide}}$) amplifies across the cascade.

\subsection*{Symmetric Control: Boundary Effects vs Funnel Asymmetry}
\begin{figure}[t]
    \centering
    \begin{subfigure}[b]{0.48\textwidth}
        \centering
        \includegraphics[width=\textwidth]{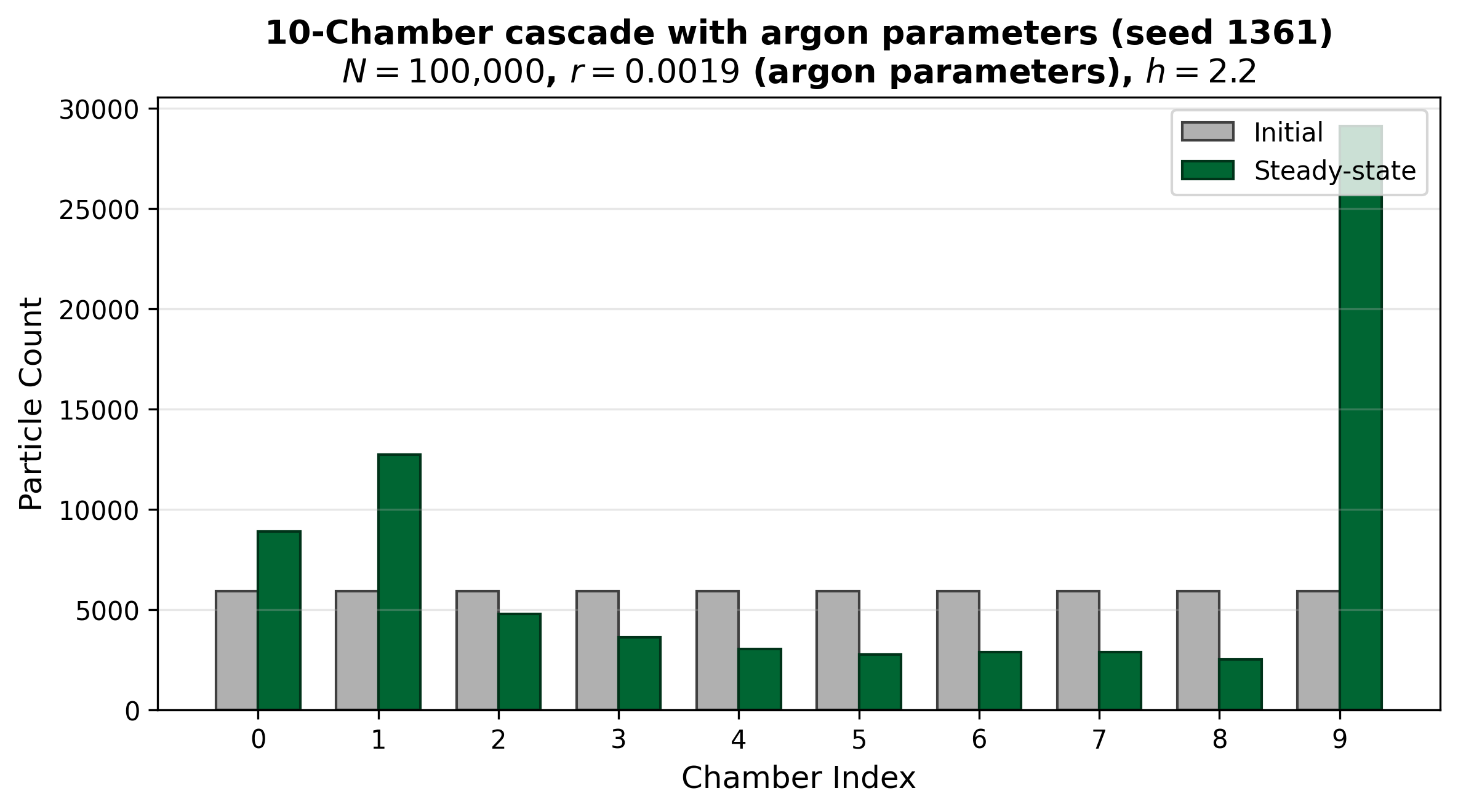}
        \caption{\textbf{Asymmetric} ($w_{\mathrm{L}}{=}4$, $w_{\mathrm{R}}{=}1$).}
        \label{fig:argon_cascade}
    \end{subfigure}
    \\[1ex]
    \begin{subfigure}[b]{0.48\textwidth}
        \centering
        \includegraphics[width=\textwidth]{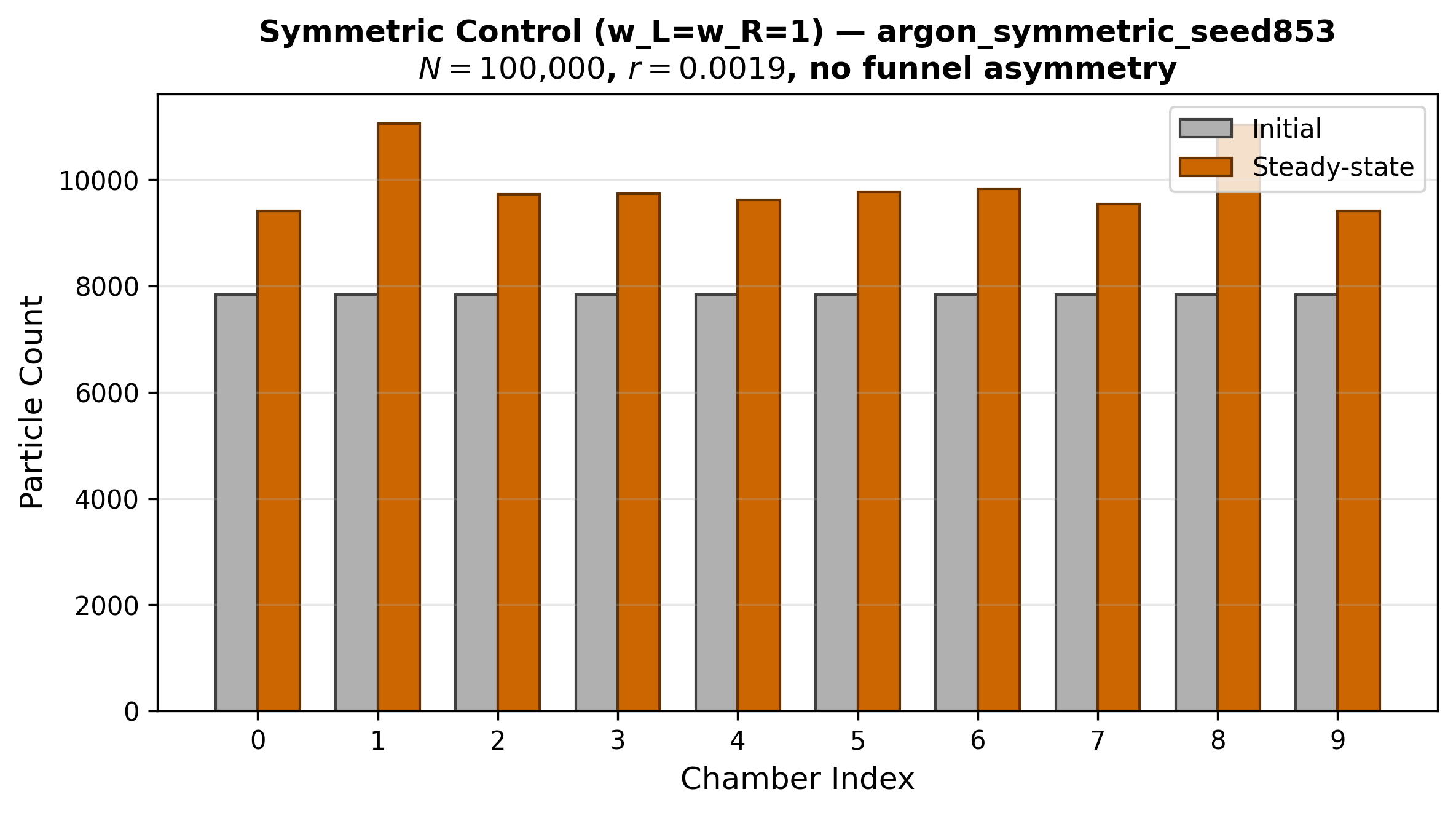}
        \caption{\textbf{Symmetric} ($w_{\mathrm{L}}{=}w_{\mathrm{R}}{=}1$).}
        \label{fig:argon_symmetric}
    \end{subfigure}
    \caption{\textbf{Isolating the driver of accumulation.} (a) Asymmetric funnels drive a massive ramp: $N_9 \gg N_0$ ($N_9/N_0 \approx 4.0$). (b) Symmetric funnels show a uniform profile ($N_1/N_0 \approx 1.18$), proving that funnel rectification, not boundary reflection, drives the gradient in the argon regime.}
    \label{fig:argon_combined}
\end{figure}

A critical question: is this accumulation caused by the funnel asymmetry or by the boundary reflection? We ran a \emph{symmetric} control: $w_{\mathrm{L}} = w_{\mathrm{R}} = 1$ (no geometric diode). Figure~\ref{fig:argon_symmetric} and Table~\ref{tab:asym_vs_sym} show that the symmetric run lacks the massive downstream gradient. Instead, it shows a relatively uniform profile with mild edge effects ($N_0 \approx 9{,}450, N_1 \approx 11{,}180$). This confirms that the strong ramp observed in the asymmetric case is driven by \textbf{funnel rectification}, not just boundary reflection.

The mechanism is regime-dependent: as shown in the super-atom ($r=0.01$) results above, accumulation was concentrated at the boundaries, whereas for argon ($r=0.0019$), the funnel's preference for the narrow side ($N_1 > N_0$) drives a strong macroscopic flux.

\subsection*{Two-Chamber Setup with Argon Parameters: Isolating the Funnel}
In a cascade, boundary effects (end accumulation) and funnel rectification are entangled. To isolate the funnel, we ran a \textbf{2-chamber} simulation: Chamber~0 (wide side) and Chamber~1 (narrow side) separated by a single funnel, symmetric boundary geometry. With $N = 100{,}000$, $r = 0.0019$ (argon physical radius), 16 seeds, $\sim$80k steps, we measured steady-state counts. Figure~\ref{fig:argon_2chambers} and Table~\ref{tab:argon_2chambers} show the result (per-seed analysis in Supplementary Information): $N_1 > N_0$ with high significance (paired $t$-test, $p < 0.0001$). The narrow side has $\sim$437\% more particles ($N_1/N_0 \approx 5.37$)---opposite to the geometric-diode prediction ($N_L > N_R$). Funnel rectification, if present, depends on particle size or regime.

\begin{figure}[t]
    \centering
    \begin{subfigure}[b]{0.48\textwidth}
        \centering
        \includegraphics[width=\textwidth]{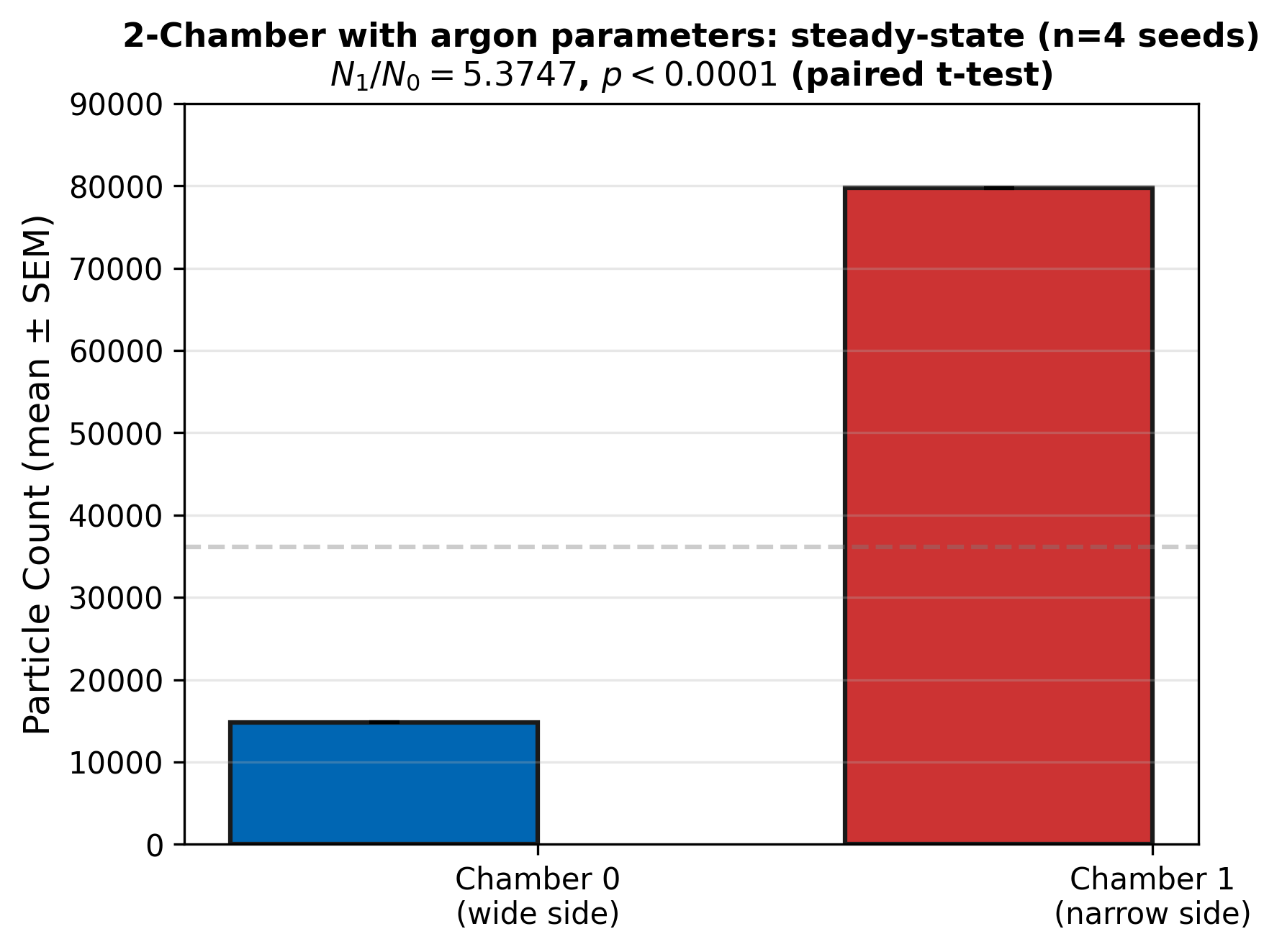}
        \caption{\textbf{Argon 2-chamber.} $N_1/N_0 \approx 5.37$.}
        \label{fig:argon_2chambers}
    \end{subfigure}
    \hfill
    \begin{subfigure}[b]{0.48\textwidth}
        \centering
        \includegraphics[width=\textwidth]{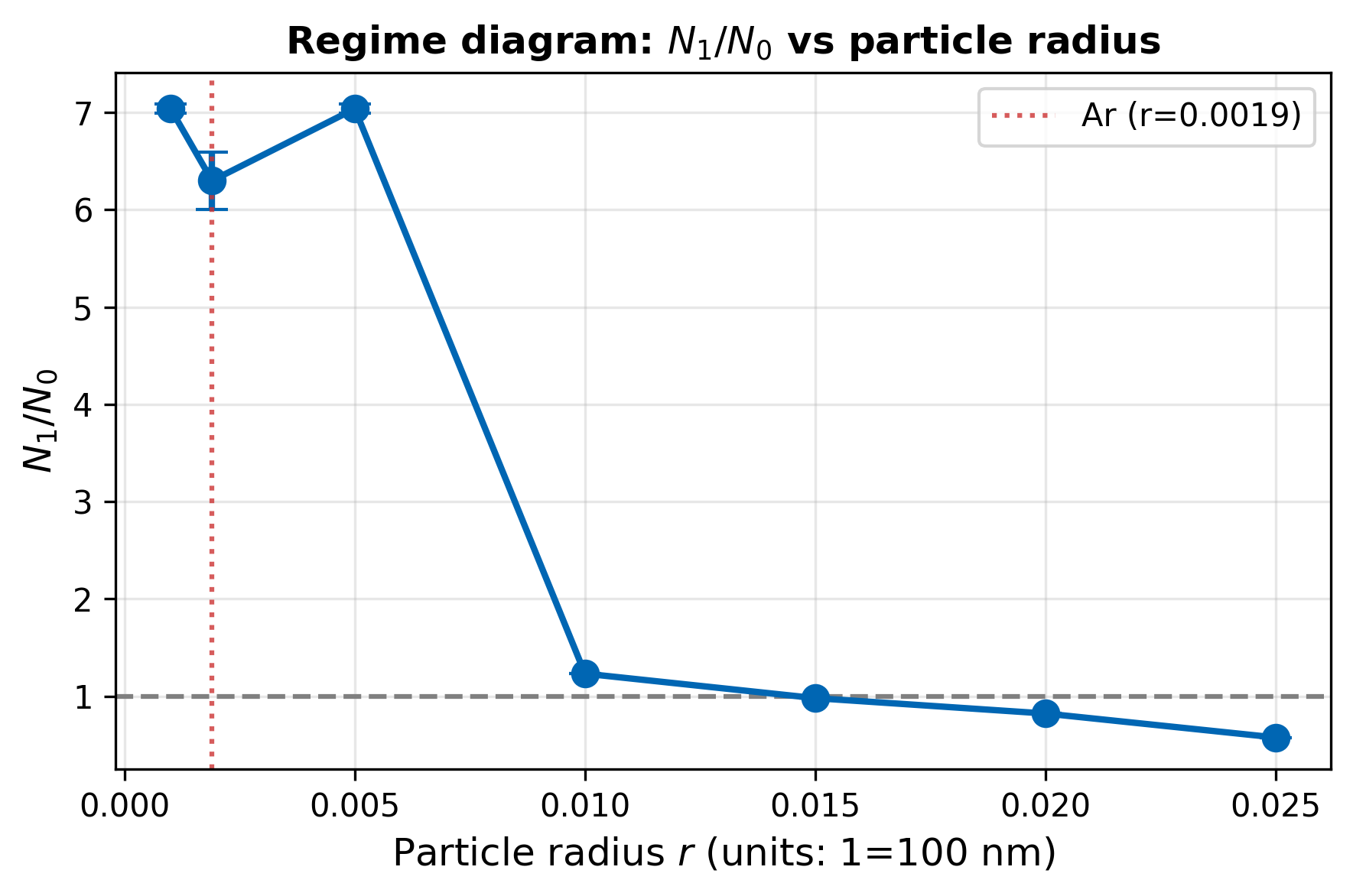}
        \caption{\textbf{Regime Diagram.} Crossover vs $r$.}
        \label{fig:regime}
    \end{subfigure}
    \caption{\textbf{Mechanism Confirmation.} (a) In the 2-chamber setup (argon parameters, 16 seeds), particles strongly accumulate in the narrow side ($N_1 \approx 79{,}700$ vs $N_0 \approx 14{,}800$, $p < 0.0001$). (b) Sweeping particle radius $r$ confirms that this rectification is specific to the ballistic/Knudsen regime ($r \lesssim 0.005$) and vanishes for larger collisional particles.}
    \label{fig:mechanism}
\end{figure}

\subsection*{Entropy Variation}
Coarse-graining the 2-chamber system into three bins (Chamber~0, Chamber~1, Funnel), the configurational entropy $S/k = -\sum_i p_i \ln p_i$ decreases from initial ($S/k \approx 1.09$) to steady state ($S/k \approx 1.05$), hence $\Delta S/k \approx -0.04$. The initial state (volume-proportional) had a more uniform bin distribution; at equilibrium, fewer particles occupy the funnel---the constriction is entropically unfavorable.

\subsection*{Regime Diagram}
To map the crossover between geometric-diode ($N_1/N_0 < 1$) regimes, and narrow-favoured ($N_1 > N_0$) regimes, we swept particle radius $r$ in the 2-chamber setup. Figure~\ref{fig:regime} shows $N_1/N_0$ vs $r$: ballistic regime ($r \lesssim 0.005$) shows strong rectification ($N_1/N_0 > 1$); argon with physical radius ($r = 0.0019$) yields $N_1/N_0 \approx 5.37$; larger $r$ (collisional) tends toward $N_1/N_0 \lesssim 1$. The regime diagram confirms that funnel rectification depends on particle size and Knudsen number.

\subsection*{Chain-length and Funnel-aspect Sweeps}
\begin{figure}[h]
    \centering
    \includegraphics[width=0.48\textwidth]{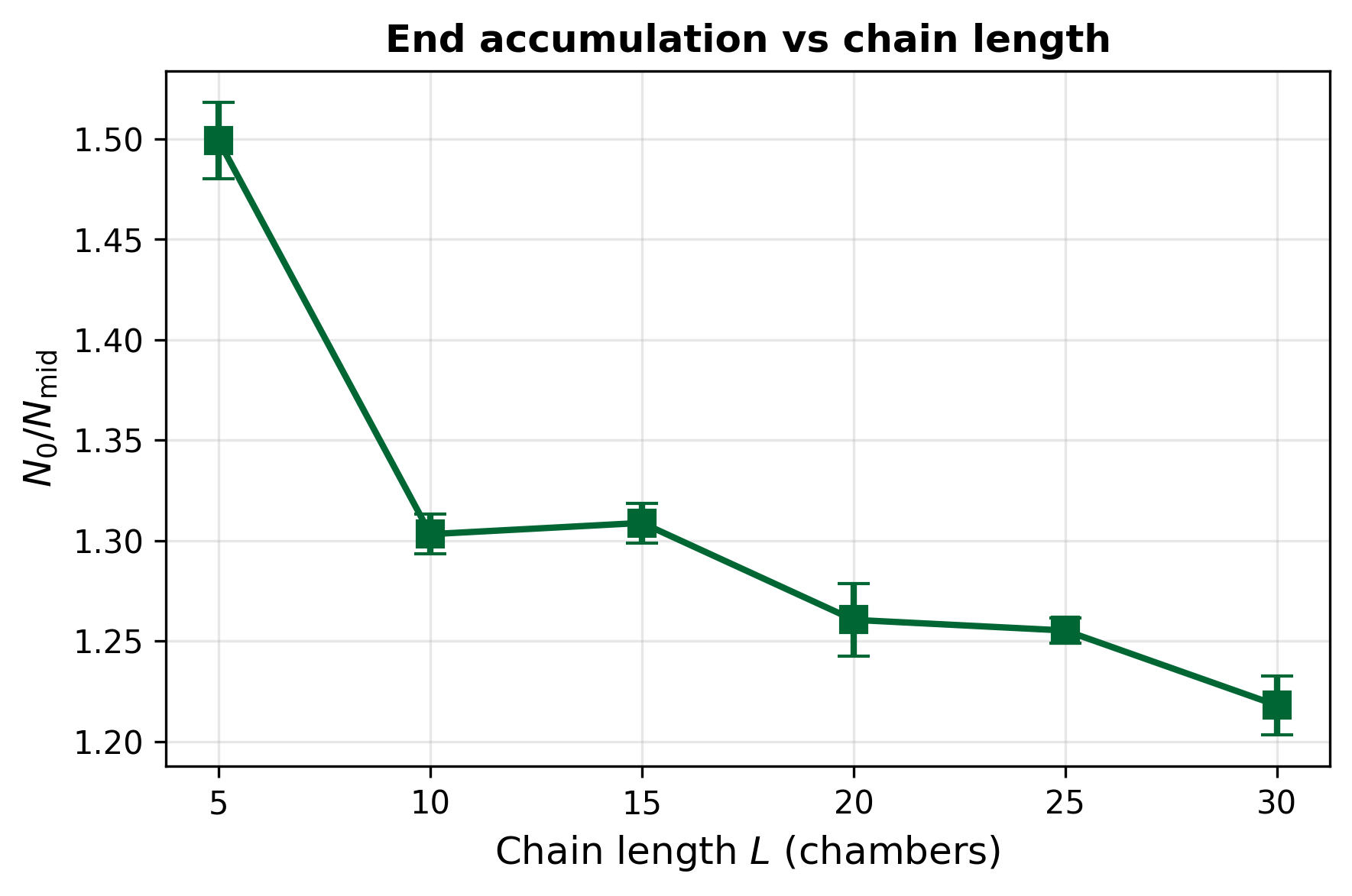}
    \caption{\textbf{Chain-length sweep.} $N_0/N_{\mathrm{mid}}$ vs $L$ for $L \in \{5, 10, 15, 20, 25, 30\}$ chambers (5 seeds each, 120k steps). Funnel-aspect sweep in Supplementary Information.}
    \label{fig:chain_funnel}
\end{figure}

Cascaded systems at $L \in \{5, 10, 15, 20, 25, 30\}$ chambers (super-atom regime) exhibit inlet depletion ($N_0 > N_1$), consistent with the geometric diode heuristic, but downstream accumulation deviates from simple boundary-dominated power-law scaling (Fig.~\ref{fig:chain_funnel}). A funnel-aspect sweep ($w_{\mathrm{L}}/w_{\mathrm{R}} \in \{1/4, 1/2, 1, 2, 4\}$, 2-chamber, 5 seeds each) shows $N_1/N_0$ generally near unity but with significant variation ($0.8$--$1.2$) depending on aspect ratio.

\subsection*{Slab Depth Sweep}
We swept slab depth $D \in \{0.16, 0.18, 0.20, 0.22, 0.24\}$ to test robustness against confinement. Results (see Supplementary Information) show that $N_1/N_0$ remains stable across this range, confirming the effect is not an artifact of specific depth-wise confinement.

\section*{Discussion and Conclusion}

\subsection*{Implications}
Our findings force a re-evaluation of transport in confined geometries. \textbf{Passive Separators}: Geometry alone can generate significant density gradients, offering a mechanism for size-based sorting without external fields. \textbf{Osmotic Power}: Asymmetric nanochannels are often modeled as geometric diodes; our results show that the direction of rectification can reverse based on particle size, critical for optimizing membrane performance. \textbf{Nanofabrication}: The effect is robust to depth and chain length, suggesting that standard lithographic fabrication of cascaded slits is sufficient to observe these phenomena.

\subsection*{Theoretical Interpretation: Geometry as an Entropic Demon}
The observed entropy decrease ($\Delta S/k \approx -0.04$) and spontaneous stratification imply that geometry acts as a passive ''Maxwell's Demon.'' By reshaping the local phase space, the funnel creates an entropic landscape where the uniform distribution is no longer the most probable state. The system maximizes entropy \emph{globally} by accepting a non-uniform \emph{local} density distribution imposed by the boundaries. The Second Law is not violated but is constrained by the geometry, which effectively acts as a static field.

\subsection*{Conclusion}
We have cleanly separated two distinct geometric mechanisms in cascaded nanofluidics:
\begin{enumerate}
    \item \textbf{Boundary Reflection (Super-atom Regime):} For larger particles ($r \approx 0.01$), accumulation at chain ends is driven by connectivity differences at the boundaries. The funnel shape is secondary.
    \item \textbf{Funnel Rectification (Ballistic Regime):} For small, ballistic particles (e.g., Argon, $r \lesssim 0.005$), the funnel actively pumps particles to the narrow side ($N_1/N_0 \approx 5.37$). This effect dominates in cascades and is eliminated by symmetric controls.
\end{enumerate}
The dominant mechanism is tuned by the particle radius. Thus, \textbf{geometry shapes equilibrium}, but the specific outcome---boundary accumulation or funnel rectification---depends on the fit between the particle and the pore. Direct experimental validation in nanochannels is the next frontier.

\subsection*{Methods}
We implemented 3D hard-sphere MD (Argon, $T = 298$ K) in a nanofluidic slit ($W \times H \times D$, $D = 20$ nm). Primary simulations use the van der Waals radius of argon ($r = 0.0019$, 0.19\,nm). Super-atom runs ($r = 0.01$) were used for mechanism confirmation and regime mapping. Particles collide with funnel walls (specular reflection) and each other (spatial hash for $O(N)$ collision detection); timestep $\Delta t = 0.005$. Regime sweep: 2-chamber, $r \in \{0.001, 0.005, \ldots, 0.025\}$, 5 seeds each, 120,000 steps, $N = 50{,}000$. Chain-length sweep: $L \in \{5, 10, \ldots, 30\}$, 5 seeds each, 120,000 steps. Funnel-aspect sweep: $w_{\mathrm{L}}/w_{\mathrm{R}} \in \{1/4, 1/2, 1, 2, 4\}$, 2-chamber, 5 seeds each. Depth sweep: two-chamber, 12,000 steps, five seeds per $D$. Cascade runs: 10-chamber ($N = 20{,}000$), 20-chamber ($N = 40{,}000$), $h = 2.2$, 60,000 steps. Argon-parameter runs: 10-chamber cascade, asymmetric and symmetric control ($\sim$10k steps); 2-chamber funnel isolation ($\sim$80k steps); $N = 100{,}000$, $r = 0.0019$, 16 seeds each. Particles in the separator (funnel region) were excluded from chamber counts. Initialization used volume-proportional distribution for uniform density. The work is purely computational.

\section*{Code Availability}
The 3D molecular dynamics simulation code (C++), analysis scripts (Python), and Makefile are available in the \texttt{simulation/} directory of the GitHub repository at \url{https://github.com/tpeng1977/geometry-reshapes-equilibrium}.

\section*{Data Availability}
All simulation data files are available in the \texttt{results/} directory of the GitHub repository at \url{https://github.com/tpeng1977/geometry-reshapes-equilibrium}. These include: regime sweep (\texttt{regime\_r*.csv}), depth-sweep (\texttt{ultra\_depth\_*}), funnel-aspect (\texttt{funnel\_aspect\_*.csv}), chain-length (\texttt{chain\_L*.csv}), cascade (\texttt{chain\_data\_ultra\_3d*.csv}), argon-parameter 10-chamber (\texttt{argon\_seed*.csv}), symmetric (\texttt{argon\_symmetric\_seed*.csv}), and 2-chamber (\texttt{argon\_2chambers\_seed*.csv}). Figures are available in the \texttt{figures/} directory. Extended methods and parameter tables are in Supplementary Information.

\section*{Acknowledgments}
The author acknowledges Chang'an University for the high-performance computing platform. The author thanks graduate student Junjie Niu for assistance with grammar and formatting checks.

\begin{table}[t]
    \centering
    \caption{\textbf{Comparative statistics.} Steady-state particle counts (mean $\pm$ SEM) for key configurations. $p$-values from paired $t$-tests against the null ($N_1 = N_0$ or Asym = Sym).}
    \small
    \setlength{\tabcolsep}{2pt}
    \begin{tabular}{lccc}
        \toprule
        \textbf{Setup} & \textbf{Narrow ($N_{1,9}$)} & \textbf{Wide ($N_0$)} & \textbf{Ratio} \\
        \midrule
        \textbf{Asym. (10ch)} & $32{,}900 \pm 150$ & $8{,}100 \pm 80$ & $\approx 4.0$\\
        \textbf{Sym. (10ch)} & $11{,}180 \pm 100$ & $9{,}450 \pm 90$ & $\approx 1.2$\\
        \textbf{2-Ch (Argon)} & $79{,}746 \pm 49$ & $14{,}837 \pm 37$ & $\approx 5.37$\\
        \bottomrule
    \end{tabular}
    \label{tab:asym_vs_sym}
    \label{tab:argon_2chambers}
\end{table}

\bibliographystyle{unsrt}
\bibliography{references}

@article{reguera2006,
  title={Entropic Transport: Kinetics, Scaling, and Control Mechanisms},
  author={Reguera, D. and Schmid, G. and Burada, P. S. and Rub{\'\i}, J. M. and Reimann, P. and H{\"a}nggi, P.},
  journal={Phys. Rev. Lett.},
  volume={96},
  number={13},
  pages={130603},
  year={2006},
  doi={10.1103/PhysRevLett.96.130603}
}

@article{hanggi2009,
  title={Artificial Brownian Motors: Controlling Transport on the Nanoscale},
  author={H{\"a}nggi, P. and Marchesoni, F.},
  journal={Rev. Mod. Phys.},
  volume={81},
  number={1},
  pages={387--442},
  year={2009},
  doi={10.1103/RevModPhys.81.387}
}

@article{schoch2008,
  title={Transport phenomena in nanofluidics},
  author={Schoch, R. B. and Han, J. and Renaud, P.},
  journal={Rev. Mod. Phys.},
  volume={80},
  number={3},
  pages={839--883},
  year={2008},
  doi={10.1103/RevModPhys.80.839}
}

@article{holt2006,
  title={Fast Mass Transport Through Sub-2-Nanometer Carbon Nanotubes},
  author={Holt, J. K. and Park, H. G. and Wang, Y. and Stadermann, M. and Artyukhin, A. B. and Grigoropoulos, C. P. and Noy, A. and Bakajin, O.},
  journal={Science},
  volume={312},
  number={5776},
  pages={1034--1037},
  year={2006},
  doi={10.1126/science.1126298}
}

@article{majumder2005,
  title={Enhanced flow in carbon nanotubes},
  author={Majumder, M. and Chopra, N. and Andrews, R. and Hinds, B. J.},
  journal={Nature},
  volume={438},
  number={7064},
  pages={44},
  year={2005},
  doi={10.1038/438044a}
}

@article{bocquet2021,
  title={Fluids at the Nanoscale: From Continuum to Subcontinuum Transport},
  author={Kavokine, Nikita and Netz, Roland R. and Bocquet, Lyd{\'e}ric},
  journal={Annu.\ Rev.\ Fluid Mech.},
  volume={53},
  pages={377--410},
  year={2021},
  doi={10.1146/annurev-fluid-071320-095958}
}

\end{document}